\documentclass[PHM, 2022]{PHMSociety}

\usepackage{graphicx}
\usepackage{amsmath}
\usepackage{listings}
\usepackage{array}
\usepackage{makecell,placeins,url}
\usepackage{caption}
\usepackage{subcaption}
\newcolumntype{L}[1]{>{\raggedright\let\newline\\\arraybackslash\hspace{0pt}}m{#1}}
\newcolumntype{C}[1]{>{\centering\let\newline\\\arraybackslash\hspace{0pt}}m{#1}}
\newcolumntype{R}[1]{>{\raggedleft\let\newline\\\arraybackslash\hspace{0pt}}m{#1}}

\begin{document}

\title{Computational Reproducibility Within Prognostics and Health Management}

\author{%
	Tim von Hahn, Chris K. Mechefske
}

\address{
	\affiliation{{}}{Department of Mechanical and Materials Engineering\\Queen’s University, Kingston, Canada}{ 
		{\email{t.vonhahn@queensu.ca}}\\ 
		{\email{chris.mechefske@queensu.ca}}
		} 
	\tabularnewline 
}

\maketitle

\phmLicenseFootnote{von Hahn}

\begin{abstract}
Scientific research frequently involves the use of computational tools and methods. Providing thorough documentation, open-source code, and data – the creation of reproducible computational research – helps others understand a researcher’s work. Here, we explore computational reproducibility, broadly, and from within the field of prognostics and health management (PHM). The adoption of reproducible computational research practices remains low across scientific disciplines and within PHM. Our text mining of more than 300 articles, from publications engaged in PHM research, showed that fewer than 1\% of researchers made both their code and data available to others. Although challenges remain, there are also clear opportunities, and personal benefits, for engaging in reproducible computational research. Highlighting an opportunity, we introduce an open-source software tool, called PyPHM, to assist PHM researchers in accessing and preprocessing common industrial datasets.  
\end{abstract}

\section{Introduction}

\emph{An article about computational science [...] is merely advertising of the scholarship. The actual scholarship is the complete software development environment and the complete set of instructions which generated the figures.}

The above statement, provocatively expressed by Buckheit and Donoho \cite{buckheit1995wavelab} paraphrases the thoughts of Jon Claebout. Claebout, a geophysicist, became an early advocate for reproducible computational research. Simply put, reproducible computational research is performed when “all details of the computation — code and data — are made conveniently available to others.” \cite{donoho2008reproducible}

Claebout’s advocacy, in the early 1990s, came at a time when computation was ascending as a means of conducting scientific research. Researchers were wrestling with how these new tools affected the dissemination of ideas. Today, computational research is ubiquitous across a multitude of fields. In addition, the paradigms of the internet, immense computational power, massive data, and open-source software, have enabled tremendous scientific advances. Yet, creating and encouraging reproducible computational research remains a challenge \cite{ince2012case}, \cite{trouble_lab_2013}.

Prognostics and health management (PHM) “is an enabling technology used to maintain the reliable, efficient, economic and safe operation of engineering equipment, systems and structures.” \cite{hu2022prognostics} PHM practitioners build these technologies with the same computational tools used across the breadth of modern science. As such, the field encounters similar challenges surrounding reproducible computational research.

The work presented here explores the topic of reproducible computational research from within the context of PHM. We ask four questions:
\begin{enumerate}
\item 	What does reproducible computational research look like? We present a practical example from PHM.
\item	What is the state of reproducible computational research? We look at findings from the broader scientific community and discuss our own findings from within PHM.
\item	What are the benefits to conducting reproducible computational research? We discuss the underappreciated personal benefits of conducting reproducible computational research. Namely, performing computationally reproducible research yields greater exposure of one’s work, stronger career opportunities, and increased personal satisfaction. 
\item	What are the challenges and opportunities for reproducible computational research? We segment the challenges and opportunities into three categories – experience, motivation, and resources. We discuss them broadly and from within PHM.
\end{enumerate}
In addition, we introduce an open-source software package, called PyPHM\footnote{PyPHM is publicly available on GitHub: \href{https://github.com/tvhahn/PyPHM}{https://github.com/tvhahn/PyPHM} }, to assist PHM practitioners in accessing and understanding public domain datasets and creating reproducible data workflows. Other fields have similar software packages, and we note that there is need of one within PHM. As such, we invite others to assist in this endeavor.

Isaac Newton wrote to his fellow scientist, Robert Hooke, that “if I have seen further, it is by standing on the shoulders of giants.” Newton then passed on his ideas through his published writings and texts. However, for future generations to stand on our shoulders we should pass on more than writings. The data and the code are also needed.

\section{Example of Reproducible Computational Research}

PHM methods can be categorized into physics based and data-driven methods. Work that uses the physics of failure, to produce a PHM application, may require the code to be available if other researchers are to reproduce it. However, in a data-driven approach, both the code and data are needed to reproduce the work. The data-driven approach will be the focus in this section. 

\begin{figure}[ht!]
    \centering
    \includegraphics[width=\columnwidth]{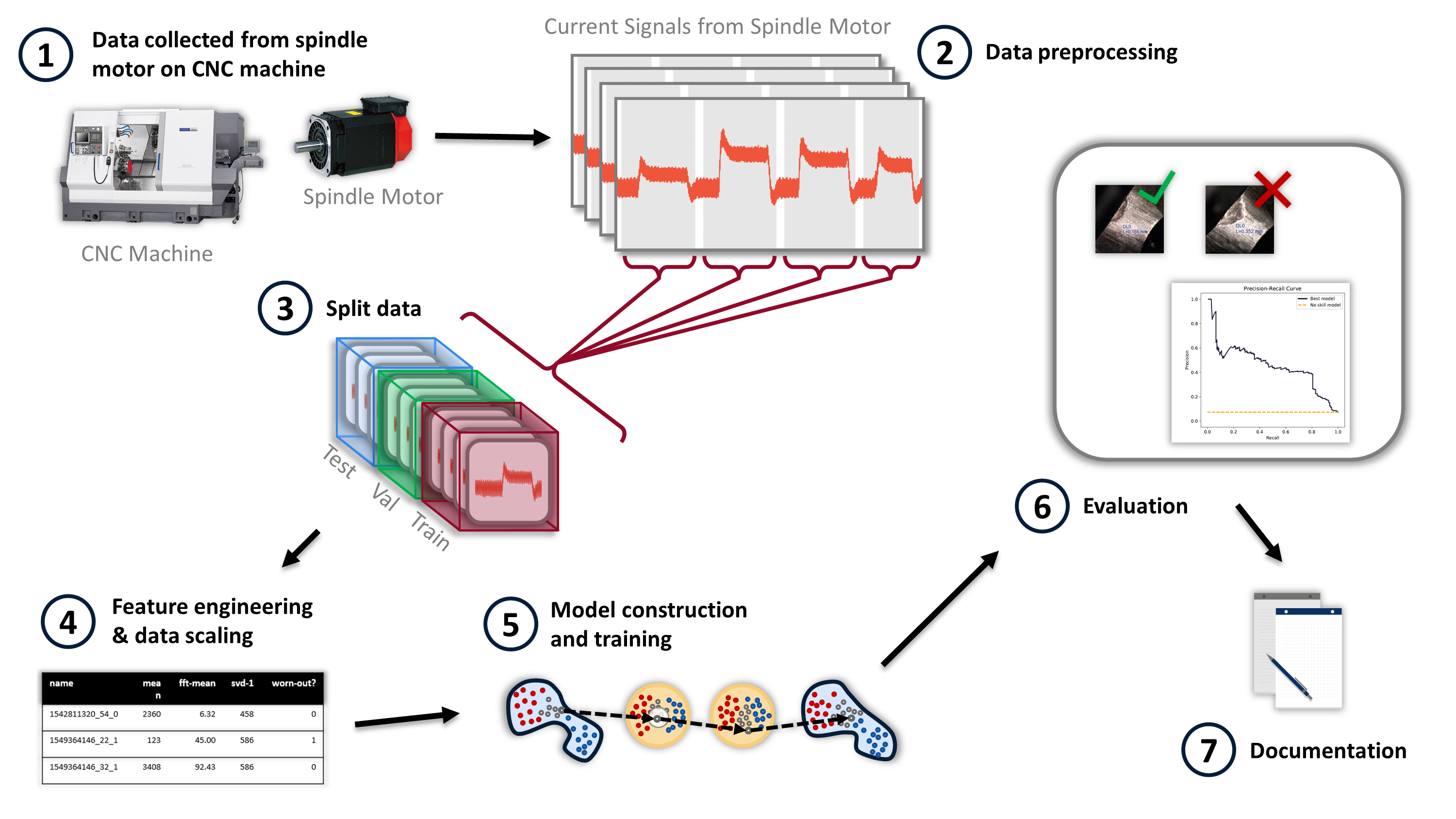}
    \caption{The simplified steps in creating a tool wear detection and prediction model for metal machining on a CNC machine.}
    \label{fig1}
\end{figure}

Figure \ref{fig1}, as an example, shows the simplified steps used to create a machine learning model for detecting tool wear on a CNC machine. The figure, combined with Table \ref{tab1}, highlights some considerations for making the work computationally reproducible.

\begin{table*}[t]\small
\caption{Creating a tool wear detection and prediction model for a CNC machine and the considerations for computational reproducibility.}
\label{tab1}
\renewcommand{\arraystretch}{1.2}
\centering
\begin{tabular}{|l|L{3cm}|l|} 
\hline
\textbf{Step} & \textbf{Description}                   & \textbf{Considerations for Computational Reproducibility}                                                                                                                                                                                                                                                                                                                                                                                                                                                                                                                                         \\ 
\hline
1             & Data Collection                        & \begin{tabular}[c]{@{}l@{}}Electrical current data is collected from the spindle motor on a CNC \\machine. Operators record when the tools are changed due to wear. ~\\Considerations: \\\begin{minipage}{12cm}\begin{itemize}
    \item 	Documentation for equipment setup and collection methodology required
\item	Meta-data for describing properties of dataset (e.g., collection frequency; time of collection; time when tools are changed; etc.) need to be provided
\item	Data should be stored on online repository accessible to public\\

\end{itemize} \end{minipage} \end{tabular}  \\ 
\hline
2             & Data
  Preprocessing                   & \begin{tabular}[c]{@{}l@{}}The raw current data is broken-up into segments. Considerations: \\\begin{minipage}{12cm}
\begin{itemize}
    \item Code and description provided showing how data is broken-up.\\
\end{itemize}
\end{minipage}\end{tabular}                                                                                                                                                                                                                                                                                                                               \\ 
\hline
3             & Data Splits                            & \begin{tabular}[c]{@{}l@{}}The data is split into training, validation, and testing data sets. \\Considerations: \\\begin{minipage}{12cm}
\begin{itemize}
    \item 	Data splits must be done before any scaling, feature engineering, or model training
\item 	Code and description of data split methodology should be provided\\

\end{itemize}
\end{minipage}\end{tabular}                                                                                                                                                                                                                \\ 
\hline
4             & Feature
  Engineering and Data Scaling & \begin{tabular}[c]{@{}l@{}}Feature engineering is conducted on the data splits. The features are then \\scaled. Considerations: \\\begin{minipage}{12cm}
\begin{itemize}
    \item 	Code and parameters used for feature engineer and scaling are made available\\
\end{itemize}
\end{minipage}\end{tabular}                                                                                                                                                                                                                                                                            \\ 
\hline
5             & Model
  Construction and Training      & \begin{tabular}[c]{@{}l@{}}Model is constructed and trained on the data that has been feature \\engineered/scaled. Considerations: \\\begin{minipage}{12cm}
\begin{itemize}
    \item 	The architecture and design decisions should be documented and made available
\item	Parameters used in model training should be recorded
\item	Hardware used is specified\\

\end{itemize}
\end{minipage}\end{tabular}                                                                                                                                                                                    \\ 
\hline
6             & Evaluation                             & \begin{tabular}[c]{@{}l@{}}The model performance is evaluated. Considerations: \\\begin{minipage}{12cm}
\begin{itemize}
    \item Metrics used to evaluate are documented
    \item	Final model saved and made available to others
    \item	Code for data visualizations is provided\\

\end{itemize}
\end{minipage}\end{tabular}                                                                                                                                                                                                                                                                      \\ 
\hline
7             & Documentation                          & \begin{tabular}[c]{@{}l@{}}Discussion of work is documented. Considerations: \\\begin{minipage}{12cm}
\begin{itemize}
    \item Paper should be made available, either as open-access, or as a preprint
\item 	Code for reproducing the results should be made available on GitHub, Gitlab, or another repository.
\item 	Software dependencies clearly defined\\

\end{itemize}
\end{minipage}\end{tabular}                                                                                                                                                                                      \\
\hline
\end{tabular}
\end{table*}

Table~\ref{tab1}, below, illustrates the complexities of reproducible computational research. Modern data-driven research contains intricate data preprocessing steps, feature engineering, and a complicated selection of parameters. Rarely can the details of this computational workflow be fully conveyed in a research paper. Consequently, the code, data, and additional documentation are needed to access this tacit knowledge. Unfortunately, many researchers only document their work through published papers. 

\section{State of Reproducible Computational Research}
The National Academies of Sciences, Engineering, and Medicine detailed the state of computational reproducibility in science in their 2019 report \cite{national2019reproducibility}. Their work showed that the lack of computational reproducibility is still a concern across science. For example, a study in computational physics demonstrated that only 6\% of articles, from 307 surveyed, make the data and code available \cite{stodden2018enabling}. 

Within the AI research domain, Gunderson et al. \cite{gundersen2018reproducible} found that fewer than 6\% of articles (out of a sample of 400) provided access to their code, and fewer than 30\% used a dataset that was publicly available. François Chollet, the creator of Keras, a popular deep learning library, laments that many deep learning papers today are “often optimized for peer review in both style and content in ways that actively hurt clarity of explanation and reliability of results.” \cite{chollet2021deep} Unfortunately, this “optimization”, as expressed by Chollet, does not encourage computational reproducibility.

Does computational reproducibility fair better within PHM? Astfalck et al. surveyed 50 PHM papers between 2000 and 2014, focusing on papers building data-driven prognostic models \cite{astfalck2016modelling}. Only eight of the papers (16\%) utilized open-source or readily available datasets, and only one paper (2\%) had the code available for inspection.

We too sought an understanding of computational reproducibility within PHM. As such, we text mined approximately 375 articles from three venues. The venues -- \href{https://www.mdpi.com/journal/energies}{Energies}, \href{https://www.sciencedirect.com/journal/mechanical-systems-and-signal-processing}{Mechanical Signals and Systems Processing (MSSP)}, and the \href{https://phm2022.phmsociety.org/}{PHM Conference} -- all feature work, to varying degrees, from within the PHM domain.

In the text mining process we searched for keywords and context that indicated whether the data or code, used in the research, was publicly available. As a comparison, we also text mined 100 computer science and electrical engineering articles from \href{https://arxiv.org/}{arXiv}, an open-access archive of scholarly articles. All the articles, from across the four venues, were randomly sampled and drawn between the years of 2015 and 2021.

\begin{figure}[htbp]
    \centering
    \includegraphics[width=\columnwidth]{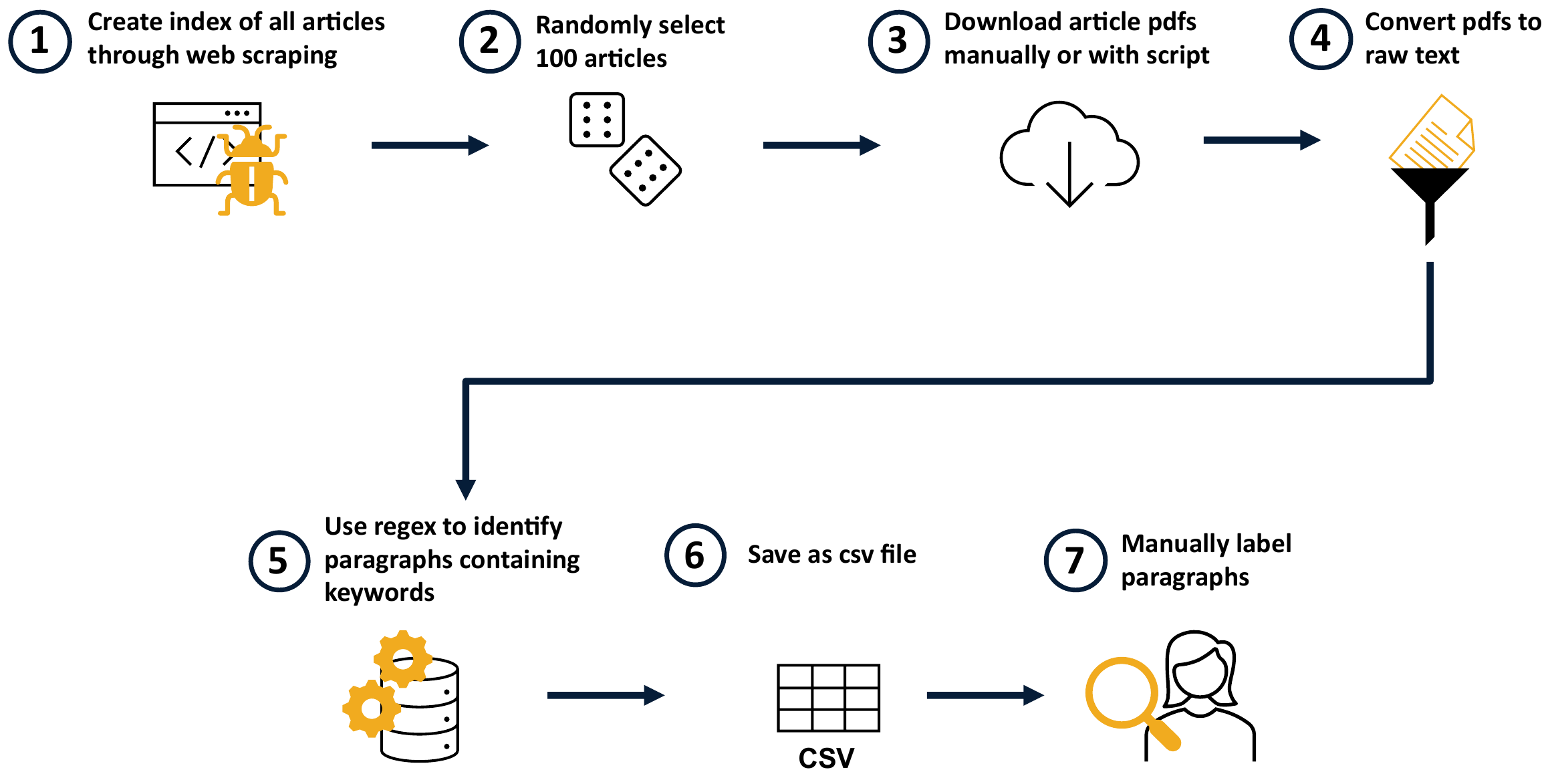}
    \caption{The steps used in the text mining process.}
    \label{text_mining}
\end{figure}

Figure~\ref{text_mining} demonstrates the text mining process for a single publication venue. The process was implemented in Python using open-source libraries. The interested reader is encouraged to visit the project page\footnote{This work, and our future text mining research, is being conducted in the open and is available on GitHub: \href{https://github.com/tvhahn/arxiv-code-search}{https://github.com/tvhahn/arxiv-code-search}} and inspect the code of this text mining process.

In the first step, as illustrated in Figure~\ref{text_mining}, web-scraping was used to build an index of all the papers published, at a specified venue, during a certain period. The web-scraping libraries \href{https://beautiful-soup-4.readthedocs.io}{Beautiful Soup} and \href{https://selenium-python.readthedocs.io/}{Selenium} were used. From there, 100 to 150 articles were randomly selected. The pdfs of the articles were then downloaded with a script, or manually, depending on the publication venue.

In step four, the raw text from the pdfs were extracted using \href{https://github.com/pdfminer/pdfminer.six}{pdfminer.six}. Regular expressions (regex) were implemented using Python’s standard library to search for keywords and short phrases, as shown in step five. Table~\ref{regex_table} demonstrates several regular expressions, amongst many, that were used in the keyword search. If a keyword match was found then the paragraph containing the keyword was saved into a csv file.

\begin{table}[ht!]\small
\caption{A sample of keywords and phrases, along with the regex code, used in the text mining process.}
\label{regex_table}
\renewcommand{\arraystretch}{1.2}
\centering
\begin{tabular}{|L{1.4cm}|L{5.8cm}|} 
\hline
\textbf{Keywords} & \textbf{Regex Code} \\ 
\hline
``used dataset'' & \lstinline!\b(used)(?:\W+\w+){0,5}?\W+(dataset)\b! \\ 
\hline
``open-source'' & \lstinline!\b(open-source|open source)\b! \\ 
\hline
``code available'' & \lstinline!\b(code)(?:\W+\w+){0,9}?\W+(available)! \\ 
\hline
\end{tabular}
\end{table}

Finally, each paragraph in the csv file was manually labelled to designate if the paragraph provided indication of public data or code. The results were then aggregated across each unique article.

\begin{figure}[ht]
    \centering
    \includegraphics[width=0.65\columnwidth]{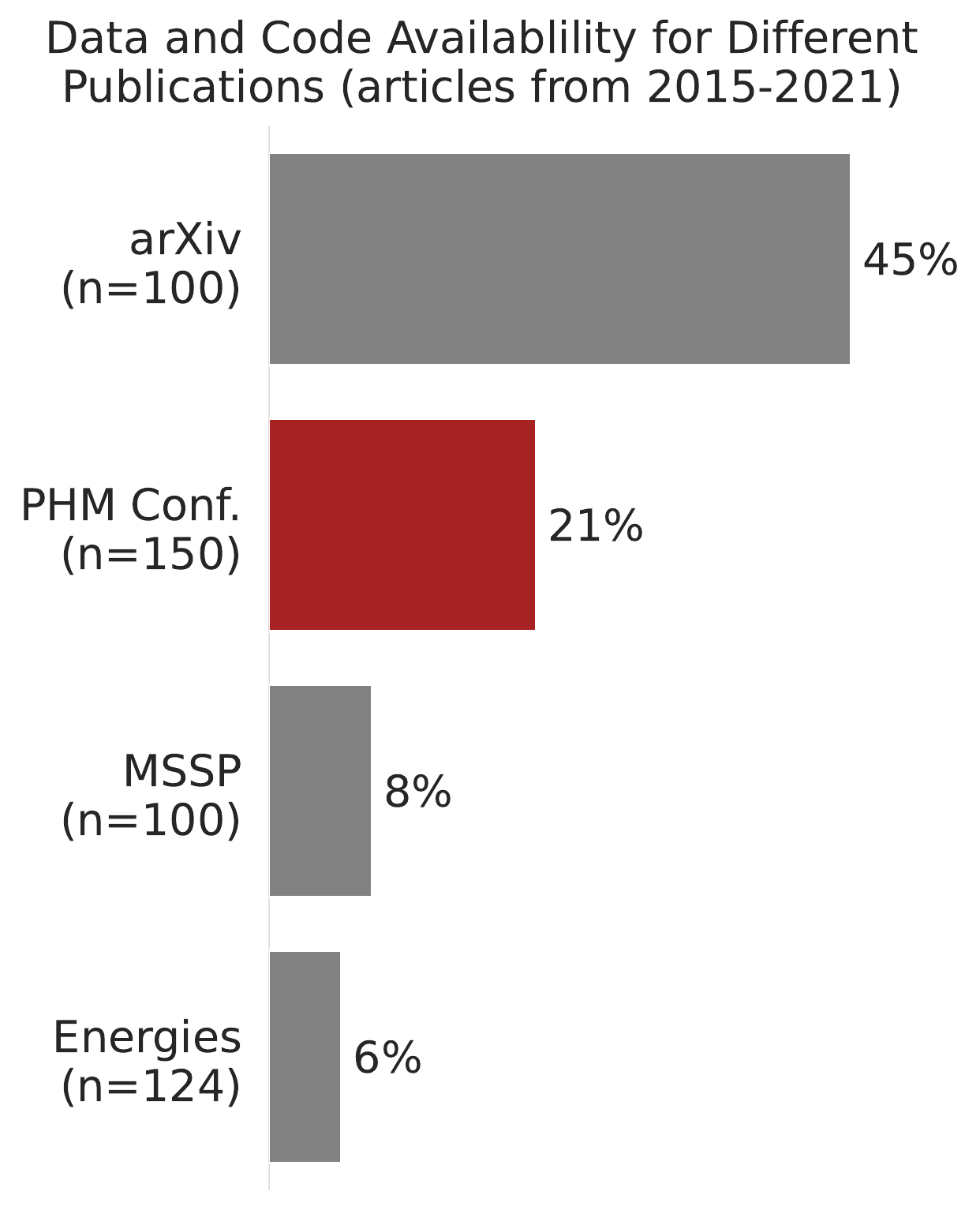}
    \caption{Percentage of articles, from various publications, that provide public access to their data or code. The articles were randomly sampled from 2015 to 2021. The sample size (e.g. n=150) for each publication venue is shown below its name.}
    \label{publishers}
\end{figure}

\begin{figure}[tp]
    \centering
    \includegraphics[width=0.8\columnwidth]{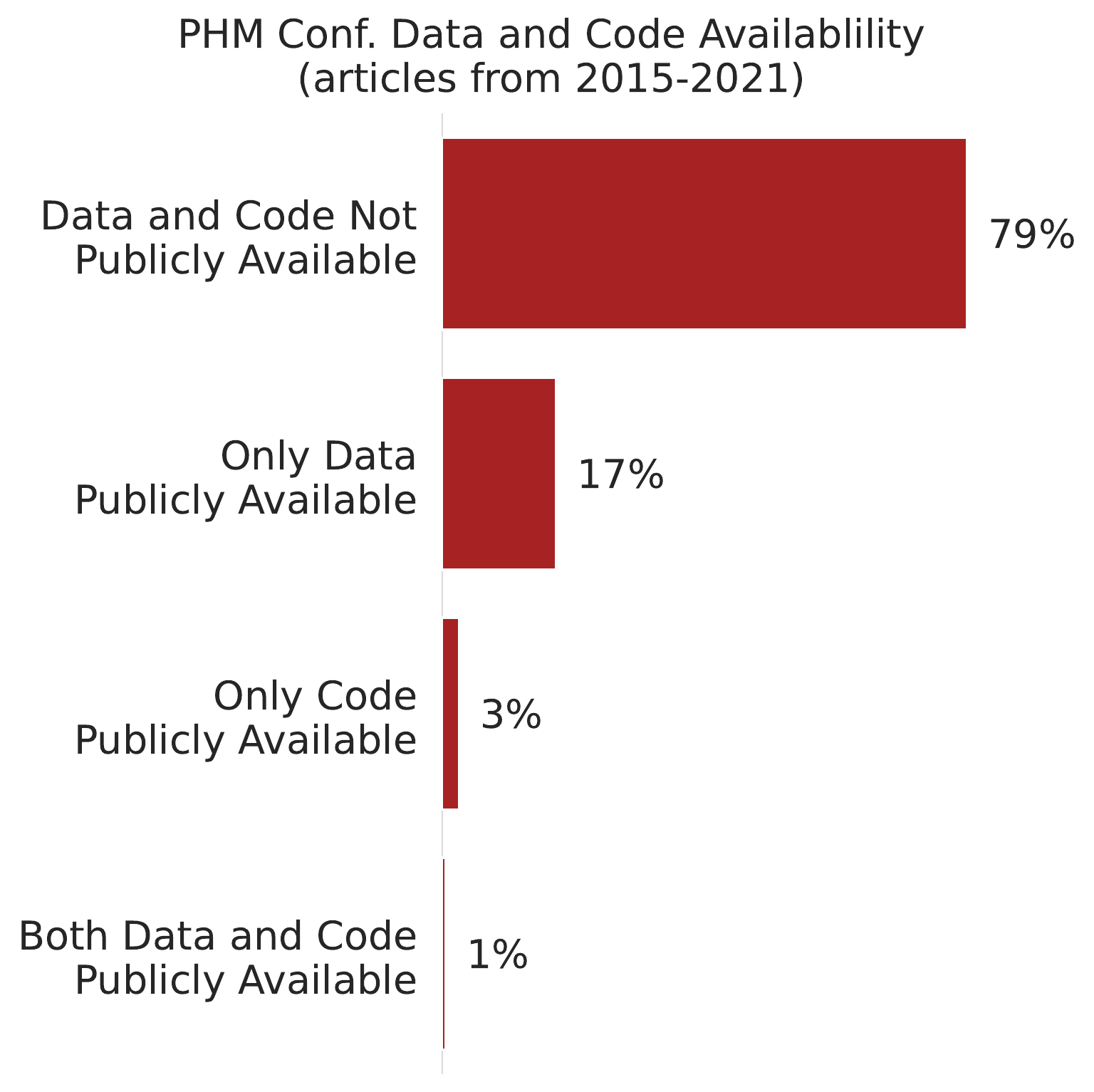}
    \caption{The distribution of data and code availability from papers published at the PHM Conferences, between 2015 and 2021. 150 articles were randomly sampled to obtain the results.}
    \label{phm_stats}
\end{figure}

From the results of the text mining, as shown in Figure~\ref{publishers}, 21\% of the technical papers from the PHM Conference provided public access to their data or code. However, only 4\% of the articles sampled from the PHM Conference, as shown in Figure~\ref{phm_stats}, provide access to the code used in their research. In general, we observed that data is much more likely to be made publicly available than the code, regardless of the publication venue. The distributions from the other publication venues are found in Figure~\ref{other_venue_pcts} in the Appendix.

Our text mining effort is ongoing as we seek a broader understanding of computational reproducibility within PHM and beyond, and we will further document these results in future publications. For now, we believe this observational study corroborates the evidence from Astfalck et al.; that is, the field of PHM suffers from similar issues of computational reproducibility as in other disciplines. 

\section{Personal Benefits of Reproducible Computational Research}
We observe that many commentators, when discussing computational reproducibility, appeal to the reader’s sense of altruism and morality. “Reproducibility is a cornerstone of the scientific method” \cite{gundersen2018reproducible} and therefore, it should be honored. In fact, we too strongly appealed to the reader’s sense of “rightness” in the introduction.

However, in this section, we highlight several benefits of computationally reproducible work that are less discussed. Rather than appealing to a sense of altruism, these appeal to the individual's self-interest. Namely, reproducible computational research can lead to increased exposure of a researcher’s work, better career opportunities, and a greater sense of satisfaction. 

\subsection{Increased Exposure}
Reproducible computational research, by its nature, requires work to be open and transparent. Most often, this necessitates that the code, data, and text are made freely available on the internet. Fortunately, this additional effort does not go unrewarded.

A significant amount of research has shown that freely releasing the code, data, and published text (either through a preprint or open-access article) leads to increased citations \shortcite{frachtenberg2022research,dorch2015data,henneken2011linking,piwowar2013data,piwowar2007sharing,colavizza2020citation,fu2019meta,christensen2019study}. In some domains, the increase was 2-fold \shortcite{wahlquist2018dissemination}. Including the code and data, alongside scientific articles, is a clear and obvious way to produce differentiated research.

\subsection{Career Opportunities}
STEM (science, technology, engineering, and mathematics) occupations are “projected to grow over two times faster than the total for all occupations in [this] decade,” according to the U.S. Bureau of Labor Statistics. Computer related occupations will produce most of this growth \cite{zilberman2021computer}. Researchers engaged in computational science will stand to benefit as their skills are increasingly in demand.

In this competitive job market, employers have begun to accept alternate credentials, as opposed to a traditional university degrees. Candidates can be hired through an intensive bootcamp program or enter a company as an apprentice. Competency can also be demonstrated through a real-world portfolio of work \cite{rainie2017future}.

Creating computationally reproducible research requires the full body of work to be accessible and understandable. Therefore, not only does reproducible computational research help the scientific community, but it also creates a strong body of work for an individual's portfolio, thus enhancing their opportunities for employment.

\subsection{Satisfaction}
The phenomenon of open-source software (OSS) consistently raises one question: why do individuals dedicate enormous amounts of their time for little economic benefit? Intrinsic motivation – a sense of internal satisfaction – is seen as a strong driver for individuals to contribute to OSS \cite{hars34working}, \cite{bitzer2007intrinsic}.

The act of producing reproducible research is like that of open-source software development. One’s work, through the code, data, and documents, is given to the world with no expectation of reward. Yet, the act of creating the work, and then selflessly sharing it with others, is internally satisfying. From the personal experience of the authors, this is one benefit of reproducible computational research that should not be ignored. 

\section{Challenges and Opportunities}
\begin{figure}[ht]
    \centering
    \includegraphics[width=1.0\columnwidth]{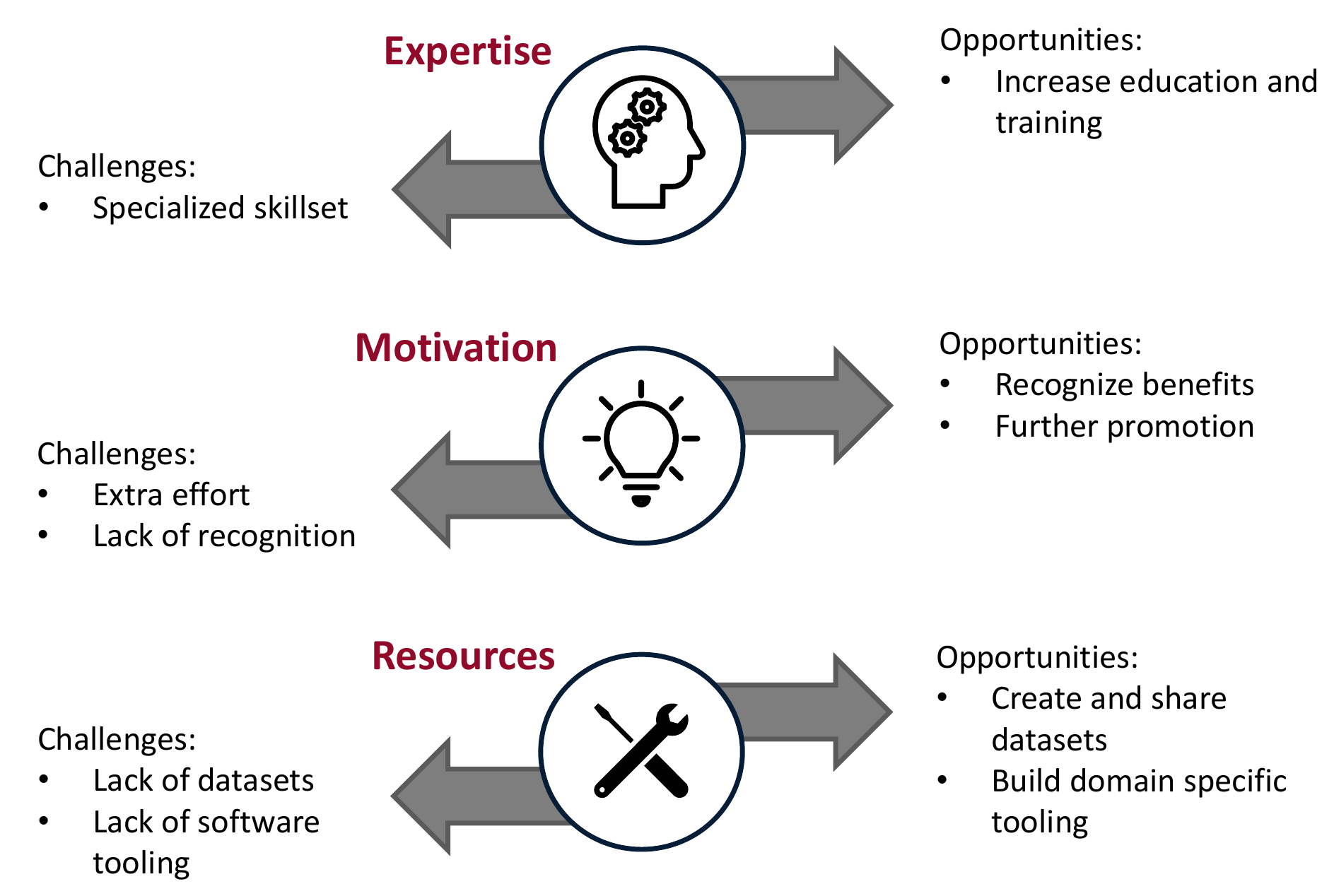}
    \caption{The three categories of challenges and opportunities for reproducible computational research.}
    \label{fig3}
\end{figure}

As discussed above, there are challenges in creating reproducible computational research. We have split these challenges, and subsequent opportunities, into three categories, as shown in Figure \ref{fig3}.

The first challenge concerns expertise. Creating reproducible computational research requires a specialized skillset that is generally covered in less depth by university curricula \cite{national2019reproducibility}. The skillset may include the use of version control, for both code and data; knowledge of containerization; or expertise in Linux, to name a few examples. 

Education efforts are being made to improve computational researcher’s expertise. Software Carpentry, a volunteer run organization, has been offering training since 1998 to improve the computational skillset of researchers \cite{wilson2014software}. Topics of computational reproducibility have been added to the curriculum, from medicine to computer science, at multiple universities \cite{neurodatascience,ucberkeleyreproducible,harvard2017reproducible}. However, we are unaware of any courses or training specific to PHM. This is an area of opportunity.

The second challenge concerns that of motivation. Reproducible computational research requires extra effort. Unfortunately, the extra effort, combined with the pressure to publish and lack of recognition, creates an impediment to reproducibility \cite{national2019reproducibility}. As discussed in Section 4, we believe that a wider recognition towards the benefits of reproducible computational research can help motivate researchers. Funding organizations, and academic journals, are also encouraging researchers to consider computational reproducibility \cite{stodden2013toward}.

Likewise, PHM specific conferences and journals should include measures to encourage computational reproducibility. As an example of this encouragement, a measure of computational reproducibility can be integrated into the peer review process. Specific recognition can also be given to papers that demonstrate computational reproducibility. Overall, a broader discussion on how to improve computational reproducibility is warranted.

Finally, the lack of domain specific resources, either in tooling or publicly available datasets, can impede computational reproducibility. The computational workflow, within a specific field, may be similar across a variety of research projects. Thus, software tools can be created to standardize these workflows. The standardization allows researchers to better grasp and more quickly reproduce each other’s work and avoids a myriad of ad-hoc approaches. The standardization also enables researchers to spend more time on higher-value tasks, such as developing novel algorithms, as opposed to preprocessing data. The software can also be coupled with open-source datasets to facilitate the comparison of results between research groups.

Table \ref{tab2}, below, lists several open-source software packages that are specific to certain domains. These software packages assist researchers in accessing datasets and reproducing computational workflows. The software, and their documentation, also assists in educating researchers on domain specific problems, techniques, and methods, and demonstrate how to implement solutions in a reproducible manner.

\begin{table}[ht!]\small
\caption{List of well-known scientific software packages, from a variety of domains, that are used to enhance computational reproducibility.}
\label{tab2}
\renewcommand{\arraystretch}{1.2}
\centering
\begin{tabular}{|L{2.2cm}|L{2cm}|L{3cm}|} 
\hline
\textbf{Software Name}               & \textbf{Domain}             & \textbf{Description}                                                                                                 \\ 
\hline
\href{https://fmriprep.org/en/stable/}{fMRIPrep} \shortcite{esteban2019fmriprep}                        & Neuroimaging                & Preprocessing pipeline for functional-MRI data                                                                       \\ 
\hline
\href{https://github.com/perone/medicaltorch}{medicaltorch} \shortcite{medicaltorchPerone}
& Medical imaging             & General package for accessing medical imaging datasets and
  standardize preprocessing methods                       \\ 
\hline
\href{https://github.com/astroML/astroML}{astroML} \shortcite{astroML}                         & Astronomy and astrophysics  & Machine learning tools and data for astronomy and astrophysics                                                       \\ 
\hline
\href{https://pytorch.org/vision/stable/index.html}{torchvision} \shortcite{NEURIPS2019_9015}                     & Computer vision             & Popular datasets,
  model architectures, and image transformations for computer vision.                              \\ 
\hline
\href{https://www.nltk.org/}{Natural Language Toolkit (NLTK)} \shortcite{Bird_Natural_Language_Processing_2009} & Natural language processing & Open-source
  modules, datasets, and tutorials supporting research and development in
  natural language processing  \\
\hline
\end{tabular}
\end{table}

As an example of this software, consider torchvision. The software allows researchers to download common computer vision datasets, apply well recognized preprocessing techniques in a standardized way, and even load already trained deep learning models. Various tutorials and examples are available to help individuals understand the functionality of the software package. 

Within PHM, there is a noted lack of high-quality, and large, datasets \cite{zhao2019deep}, \cite{wang2021recent}. The lack of these datasets may be due to the proprietary nature of industrial data or the poor understanding of their need within the PHM research community. We encourage others to freely share their PHM datasets. 

The field of PHM, to the knowledge of the authors, also lacks an open-source software package, like those found in Table~\ref{tab2}. Such a tool would enable the easy access to PHM datasets and the implementation of computational workflows. We see this as an opportunity, and as such, we have begun the process of building such a software tool, discussed below.

\section{An Open-Source Software Tool for PHM Datasets}
Currently, PHM datasets are spread-out across the internet and require users to both find the data and then manually download it. Furthermore, users must implement their own data preprocessing steps, which can be a time-consuming endeavor. The open-source software tool being developed, called PyPHM, will enable PHM practitioners to easily source, download, and preprocess publicly available PHM datasets in only a few lines of code. Researchers can use the preprocessed data from PyPHM, for example, for feature engineering or machine learning experiments.

\begin{figure}[t]
    \centering
    \includegraphics[width=.97\columnwidth]{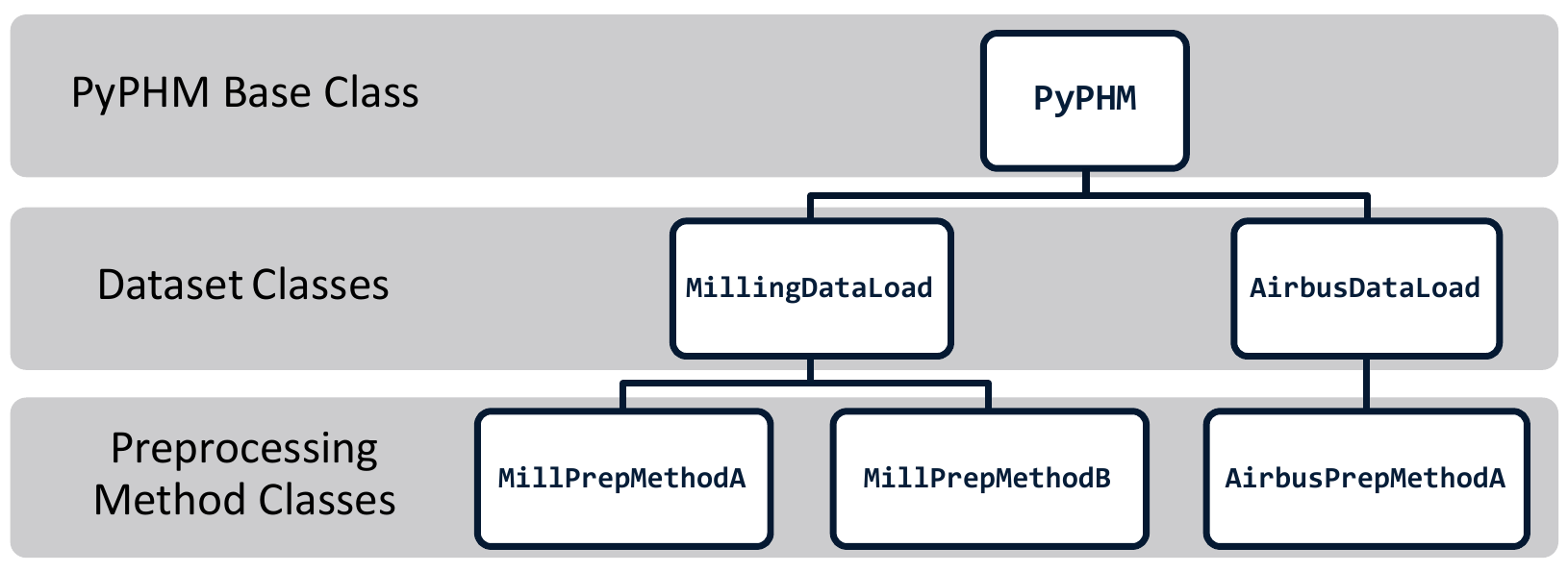}
    \caption{Class hierarchy for the PyPHM software package. The PyPHM base class implements functionality that extends across all datasets. The dataset classes implement functionality for individual datasets (the UC-Berkeley milling dataset and the Airbus helicopter dataset are shown). Finally, individual data preparation workflows are implemented for each dataset. Different preparation methods can be implemented, with each method mimicking a common workflow demonstrated by others.}
    \label{fig4}
\end{figure}

Figure \ref{fig4} illustrates the class hierarchy of the PyPHM software package. Specific datasets are accessible by their own class (beneath the base PyPHM class). The dataset class allows the downloading and extraction of the dataset, along with general preprocessing methods. Finally, simple data preprocessing methods, such as windowing, are constructed in their own preprocessing method classes. PyPHM is implemented in Python and relies on common open-source libraries like NumPy \cite{harris2020array} and SciPy \cite{2020SciPy-NMeth}.

Currently, there are three datasets implemented in PyPHM: the UC-Berkeley Milling Dataset \cite{agogino2007milling}, the IMS Bearing Dataset \cite{lee2007bearing}, and the Airbus Helicopter Accelerometer Dataset \cite{garcia2021temporal}. More will be implemented in the future.

PyPHM seeks to be a domain specific resource, within PHM, that can assist researchers in conducting reproducible computational research. We highlight three challenges, as noted in Section 5, that PyPHM seeks to specifically addresses.

\begin{enumerate}
\item 	\emph{Challenge of expertise}: PyPHM abstracts away the complexity of downloading, manipulating, and preprocessing PHM datasets. Thus, a broader audience can engage with PHM datasets, and do so in a way that can be readily reproduced by others. In addition, PyPHM is built upon common open-source tooling. Individuals can educate themselves on these tools, from a PHM context, through PyPHM’s documentation and examples.
\item	\emph{Challenge of extra effort}: PyPHM allows individuals to quickly implement a standardized workflow that other researchers have used. This saves time, reduces effort, and prevents the implementation of ad-hoc workflows that are difficult for others to reproduce. 
\item	\emph{Challenge due to lack of data}: Currently, PHM datasets are dispersed across the internet. PyPHM can act as an index for these PHM datasets and a central location to access them. PyPHM can also help individuals share and explore under-utilized PHM datasets once more datasets are added.
\end{enumerate}

PyPHM is in active development. We welcome feedback and contributions to this nascent open-source software project.

\section{Conclusion}
Computation is used across the breadth of science, and certainly within PHM. However, creating reproducible computational research remains a challenge due to the expertise required, motivational challenges, and lack of domain specific resources. Our survey of more than 300 articles, from publications engaged in PHM research, demonstrates that most researchers within PHM do not provide access to their data or code.

Despite these challenges, there are clear motivations and opportunities for improving reproducible computational research. In this work, we have highlighted three of the personal benefits of conducting reproducible computational research. Namely, creating reproducible computational research can increase a researcher’s exposure, improve career opportunities, and increase one’s sense of satisfaction. 

Furthermore, we have identified a need for an open-source software package to assist PHM researchers in accessing and preprocessing common PHM datasets. As such, we have created PyPHM, and we encourage others to assist in our efforts, either through contributions or suggestions.

\bibliographystyle{apacite}
\bibliography{sources}

\pagebreak
\onecolumn
\section*{Appendix} \label{appendix}


\begin{figure}[h]
     \centering
     \begin{subfigure}[b]{0.3\textwidth}
         \centering
         \includegraphics[width=\textwidth]{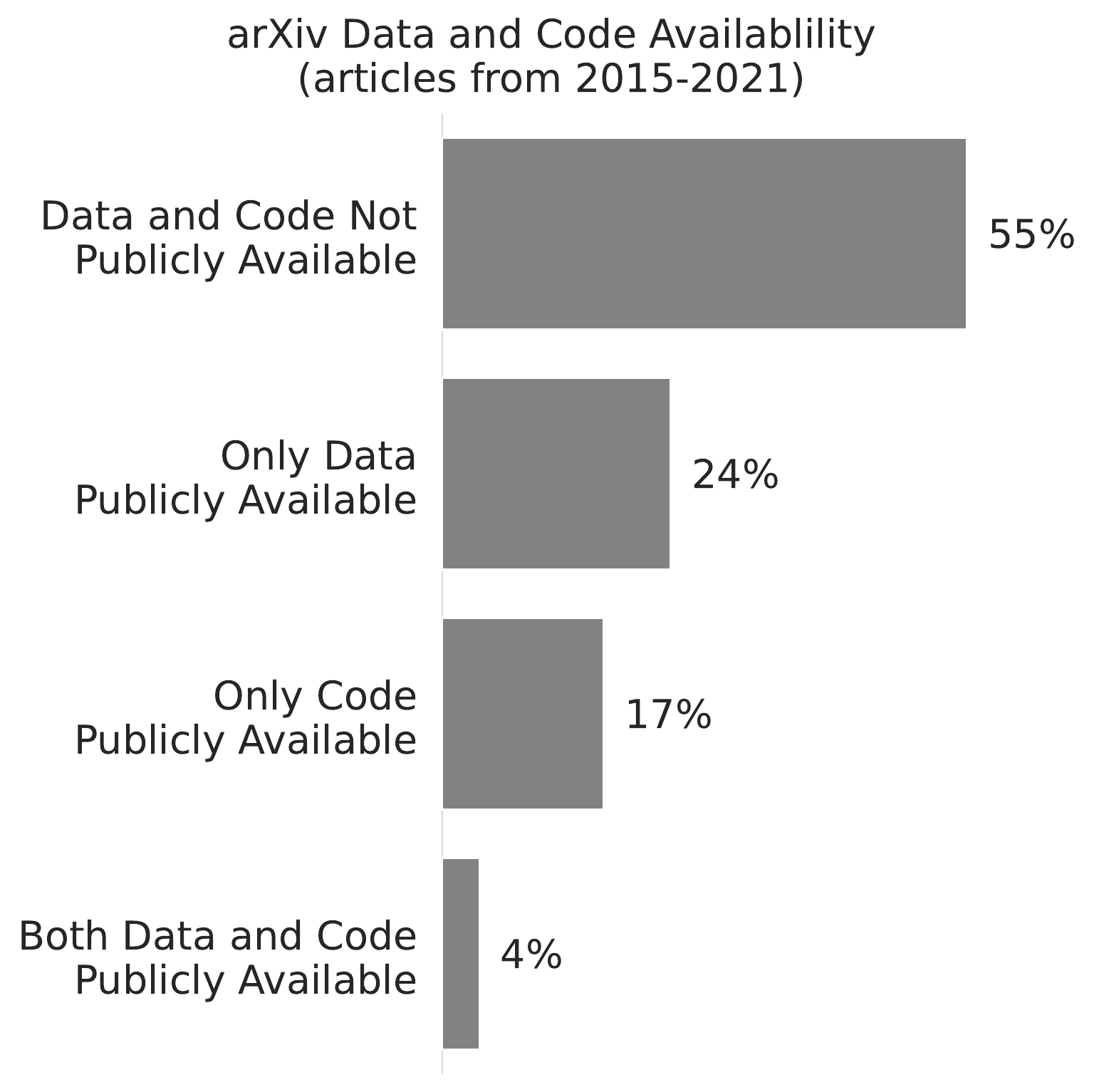}
     \end{subfigure}
     \hfill
     \begin{subfigure}[b]{0.3\textwidth}
         \centering
         \includegraphics[width=\textwidth]{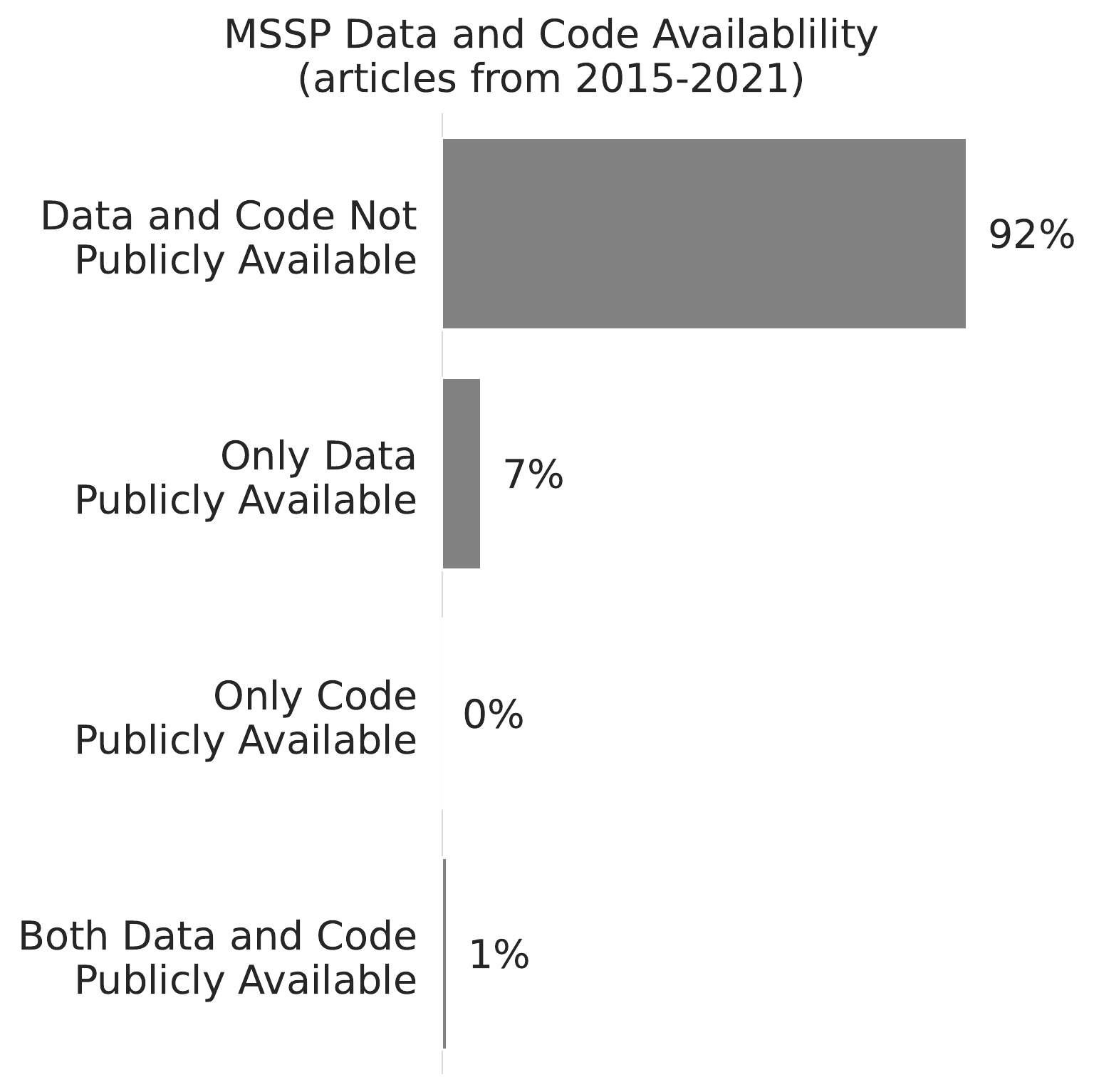}
     \end{subfigure}
     \hfill
     \begin{subfigure}[b]{0.3\textwidth}
         \centering
         \includegraphics[width=\textwidth]{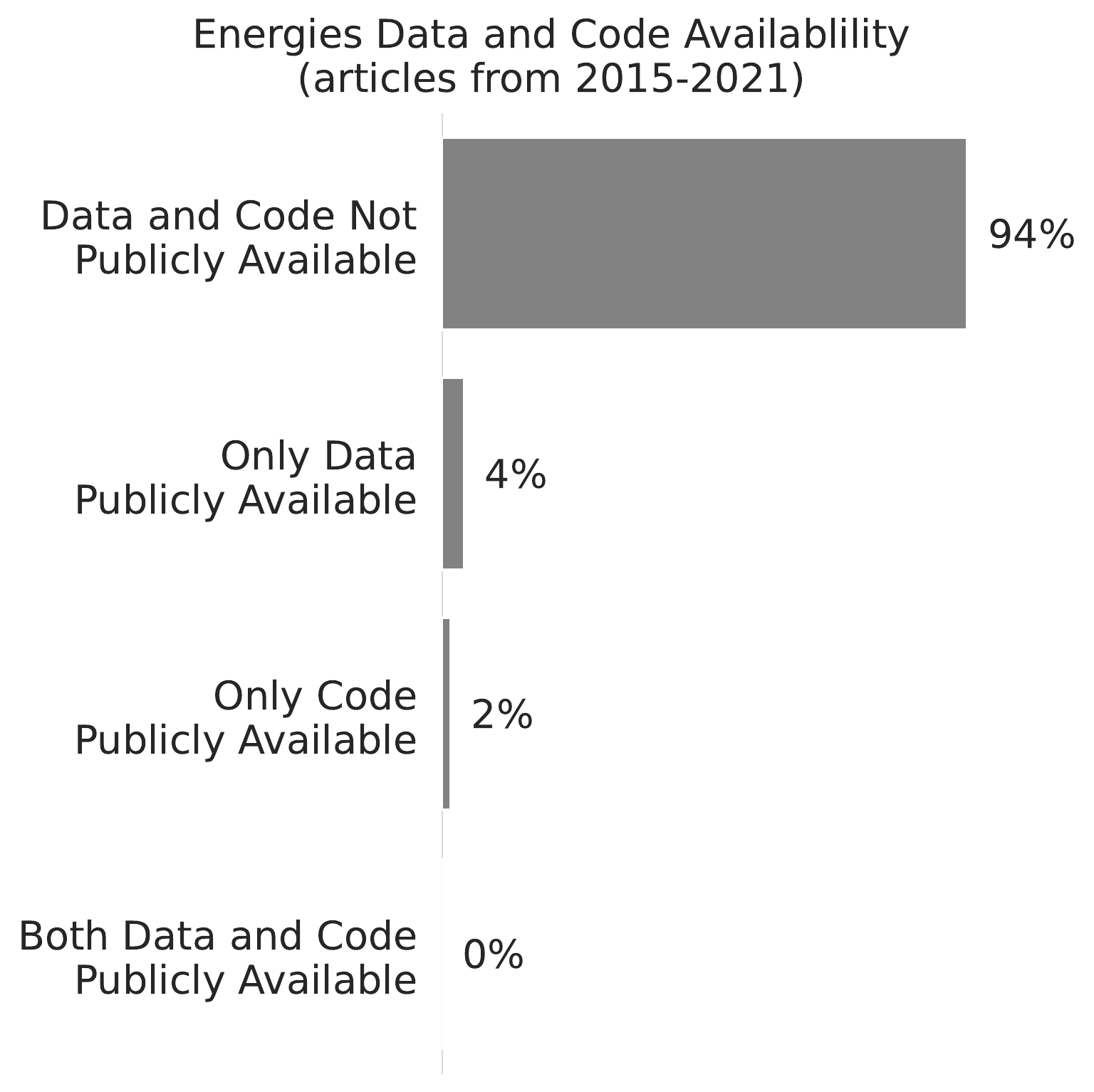}
     \end{subfigure}
        \caption{The distribution of data and code availability from papers at the arXiv, MSSP, and Energies venues. Articles were sampled between 2015 and 2021.}
        \label{other_venue_pcts}
\end{figure}

\end{document}